\documentclass[superscriptaddress,twocolumn,pra,showpacs]{revtex4}

\usepackage{amsmath,amssymb}
\usepackage{graphics}
\usepackage{epsfig}

\begin{document}

\title{Ultracold atom-molecule collisions with fermionic atoms}
\author{J. P. D'Incao}
\affiliation{Department of Physics and JILA, University of Colorado,  
Boulder, Colorado 80309-0440, USA}
\author{B. D. Esry}
\affiliation{Department of Physics, Kansas State University, Manhattan,
Kansas 66506, USA}
\author{Chris H. Greene}
\affiliation{Department of Physics and JILA, University of Colorado,  
Boulder, Colorado 80309-0440, USA}

\begin{abstract}
Elastic and inelastic properties of 
weakly bound $s$- and $p$-wave molecules of fermionic atoms that collide with a third atom are investigated. 
Analysis of calculated collisional properties of $s$-wave dimers of fermions in different spin states 
permit us to compare and highlight the physical mechanisms that determine the stability of
$s$-wave and $p$-wave molecules. 
In contrast to $s$-wave molecules, the collisional properties of $p$-wave molecules are 
found to be largely insensitive to variations of the $p$-wave scattering length and that these collisions will usually result in 
short molecular lifetimes. 
We also discuss the importance of this result for both theories and experiments involving degenerate Fermi gases.
\end{abstract} 

\maketitle 

\section{Introduction}

In the past decade, the advent of magnetic field control of interatomic interactions near a Feshbach resonance has rapidly 
sparked both experimental capabilities and theoretical understanding in the field of ultracold quantum gases. Using $s$-wave 
Feshbach resonances in an atomic Fermi gas with two different spin states \cite{TwoSpinFermiGas}, fundamental problems 
such as the crossover between Bose-Einstein condensation  and the Bardeen-Cooper-Schrieffer-type  superfluidity \cite{BECBCS} 
have become accessible experimentally, allowing  tests of fundamental theories and studies of novel phenomena (see \cite{TwoSpinFermiGas} 
and references within). Recently, the observation of $p$-wave Feshbach resonances in spin-polarized Fermi gases 
\cite{PwaveRes} instigated theoretical studies that predicted a variety of novel many-body phenomena \cite{PwaveTheory}. 
Moreover, the recent experimental observation of $p$-wave molecules \cite{PwaveMolecules} in a $^{40}$K Fermi gas has 
provided a starting point for further investigations of novel many-body phenomena in the strongly interacting regime 
\cite{PwaveTheory}. The experimental realization of the theoretical predictions will hinge on the lifetime and stability 
of $p$-wave molecules, since, e.g., the long lifetime predicted \cite{Petrov} and found \cite{TwoSpinFermiGas,SwaveMolecules}
for $s$-wave molecules was one of the key ingredients that facilitated the observation of fermionic superfluid behavior in dilute 
gases. In Ref.~\cite{PwaveMolecules}, though, it was found that $p$-wave molecules are short-lived which limits the possible 
types of states that can be created for $p$-wave molecular condensates.

In this paper, we explore the physics of three-body collisions involving both $s$-wave and $p$-wave molecules with the goal of understanding 
what determines their different levels of stability. For both $s$- and $p$-wave cases, the dominant contribution to the inelastic atom-molecule rate 
coefficient at sufficiently low collision energies has $s$-wave character and therefore does not vanish, even at zero temperature. 
Such processes thus play an important role for atom-molecule mixtures near a $p$-wave Feshbach resonance. Our results indicate 
that many of the characteristics of $p$-wave $^{40}$K$_2$ molecules are likely to be shared by other atomic species. This expectation 
is based on the universal nature of our results that are discussed below. We have also found that, near a $p$-wave Feshbach resonance, $s$-wave 
atom-molecule collisions are likely to be insensitive to the presence of the Feshbach resonance, in contrast to the $s$-wave 
case. However, our results indicate that while inelastic $p$-wave atom-molecule collisions are suppressed, elastic $p$-wave atom-molecule 
collisions are enhanced near a $p$-wave resonance. We then briefly discuss the three-body parameters that might be important for 
understanding many-body aspects in degenerate Fermi gases with resonant $p$-wave interactions.
The primary goal of the present study is to discuss calculations carried out for both $s$- and $p$-wave molecules that highlight the main 
mechanisms that cause their lifetimes to be so different.

Near a two-body Feshbach resonance, the physics that determines the collisional properties of $s$-wave 
molecules is closely related to what has been called Efimov physics \cite{Efimov,DIncao}. Efimov physics 
occurs when the two-body $s$-wave scattering length, $a_{s}=-\lim_{k \rightarrow 0}\tan{\delta_{s}}/k$
with $\delta_{s}$ the phase shift and $k$ the wave number, is abnormally large compared to the characteristic 
range $r_{0}$ of the interatomic interactions. 
For $s$-wave molecules composed of fermions in different spin states,  say $FF'$,  
Efimov physics predicts a universal repulsive effective interaction for $FF'+F$ collisions which is responsible for the $a_s^{-3.33}$ suppression of 
atomic and molecular losses  \cite{Petrov,DIncao}.
For $FF'+F$ collisions, Efimov physics also predicts that the atom-molecule 
scattering length should be $a_{ad}\approx 1.2a_{s}$ \cite{Petrov}, which ensures a strong 
atom-molecule interaction in the many-body quantum gas, and which provides efficient 
evaporative cooling near a Feshbach resonance.  A positive $a_{ad}$ can also protect the atom-molecule mixture against collapse.

However, as has been 
shown experimentally \cite{PwaveMolecules}, near a $p$-wave Feshbach resonance the situation differs strikingly and 
$p$-wave molecules composed of spin-polarized fermions in a single internal substate, say $FF$, tend to have short lifetimes.
Only recently has atom-molecule scattering for large $p$-wave scattering 
lengths, $a_{p}=\lim_{k \rightarrow 0}(-\tan{\delta_{p}}/k^3)^{1/3}$, been studied \cite{Castin,Gurarie}. 
(Note that the $p$-wave scattering is sometimes characterized equivalently by the ``$p$-wave scattering volume'' 
$V_p \equiv a_p^3$ \cite{OurFermion,Castin}.) 
Three-body recombination of spin-stretched fermions received earlier attention \cite{PwaveRes,OurFermion} where it
was shown that recombination rate could be significant near a $p$-wave resonance despite the Fermi statistics suppression.
More recently, the recombination channel in the $FFF$ system was studied using a zero-range $p$-wave pseudopotential interaction
\cite{MacekPRL}.
There, the existence of a large number 
of weakly bound three-body states analogous to the Efimov states for bosons \cite{Efimov} was proposed. 
In our calculations, we found similar three-body states at the simplest level of approximation of the hyperspherical adiabatic representation. 
They disappear, however, once nonadiabatic effects are included, which 
leads us to speculate
that these 
nonadiabatic corrections might also eliminate the three-body states found in Ref.~\cite{MacekPRL}. Moreover, it is 
not clear whether the energy dependence neglected in their $p$-wave pseudopotentials \cite{PwavePseudoPot}
affects the universality of the results predicted in Ref.~\cite{MacekPRL}. Our results presented here do not suffer from such limitations 
because our two-body finite range model potential intrinsically includes such corrections. 

In this paper we focus on atom-molecule scattering properties and find that both elastic and inelastic $s$-wave $FF+F$ processes 
are largely and surprisingly insensitive to variations of $a_p$ and potentially explain the loss rates observed in Ref.~\cite{PwaveMolecules}. 
This result agrees with the qualitative arguments in Ref.~\cite{Castin}.
We trace the insensivity of the atom-molecule losses to the weakly attractive effective atom-molecule 
interaction and to the fact that the size of a $p$-wave molecule is mainly determined by $r_{0}$ and therefore does not change as $a_{p}$ varies, 
in contrast to $s$-wave molecules whose size is directly related to $a_{s}$.
Our results, therefore, indicate that the insensitivity to variations of $a_{p}$ of both elastic and inelastic $FF+F$ processes is likely 
to be a universal property of $p$-wave molecules. Based on these results we also speculate on which few-body parameters are important 
for many-body theories for spin-polarized Fermi gases.

\section{Theoretical background}

We have extracted the three-body elastic and inelastic collisional properties
of $s$- and $p$-wave molecules from numerically converged solutions of the three-body Schr\"odinger equation carried out in the 
adiabatic hyperspherical representation \cite{Hyperspherical}, using model finite-range two-body interactions. This representation offers a simple, 
unifying picture from which we can quantitatively determine and then qualitatively interpret the origin of the significantly different collisional 
properties of $s$- and $p$-wave molecules.

In the adiabatic hyperspherical representation, the three-body system is described in terms
of the hyperradius $R$, which gives the overall size of the system, and a set of five hyperangles \cite{Hyperspherical}, 
which mainly describe the interparticle correlations. In this representation the Schr{\"o}dinger equation reduces to a system of 
coupled ordinary differential equations given (in atomic units) by, 
\begin{eqnarray}
&&\left[-\frac{1}{2\mu}\frac{d^2}{dR^2}+W_{\nu}(R)\right]F_{\nu}(R) \nonumber \\
&&~~~~~~~~~~+\sum_{\nu'\neq\nu} V_{\nu\nu'}(R) F_{\nu'}(R)=E F_\nu(R),
\label{SchrEq}
\end{eqnarray}
\noindent
where $\mu=m/\sqrt{3}$ is the three-body reduced mass for three identical particles ($m$ being the atomic mass), 
$E$ is the total energy, $F_{\nu}$ is the hyperradial wave function, and $\nu$ is a collective 
index that represents all quantum numbers necessary to label each channel.
In the above expression, $V_{\nu\nu'}$ are the nonadiabatic couplings  calculated in terms of
the $R$ derivative of the hyperangular solutions \cite{Hyperspherical}. In Eq.~(\ref{SchrEq}), the nonadiabatic couplings 
$V_{\nu\nu'}$ drive transitions between different channels, represented by the effective potentials $W_{\nu}$.

In our calculations we have used the two-body model interaction  
\begin{equation}
v(r_{ij})=-D{\rm sech}^2(r_{ij}/r_0), 
\end{equation}
\noindent
where $r_{ij}$ is
the interatomic separation. We have taken $r_{0}$ to be the van der Waals length \cite{Julienne} for $^{40}$K atoms, 
$r_{0}=\frac{1}{2}(m C_{6})^{1/4}\approx 65$~a.u., and $m$ to be the $^{40}$K atomic mass. 
We varied $a_{s}$ and $a_{p}$ by changing the potential depth $D$ around the vicinity of the value that produces a weakly bound state of
$s$- and $p$-wave character, respectively. 
For $FFF'$ systems we adjusted $D$ to produce two deeply bound $FF'$ molecules (one $s$-wave and one $p$-wave) 
in addition to the weakly bound $s$-wave molecule, and we neglected $FF$ interactions.  For $FFF$ systems, we adjusted $D$ to produce one
deeply bound $p$-wave molecule and one weakly-bound $p$-wave molecule.
In a more realistic model for $p$-wave interactions near a Feshbach resonance, however,
one needs to consider the magnetic dipole-dipole interaction which treats the resonance in terms of its angular momentum projection 
$m_{p}=0$ and $m_{p}=\pm1$ \cite{Multiplet}. Our model, which does not include dipole-dipole interactions, is applicable to 
$m_{p}=0$ and $|m_{p}|=1$ resonances individually and does not include coupling between $m_{p}=0$ and $|m_{p}|=1$ molecules.
\begin{figure}[htbp]
\includegraphics[width=3.40in,angle=0,clip=true]{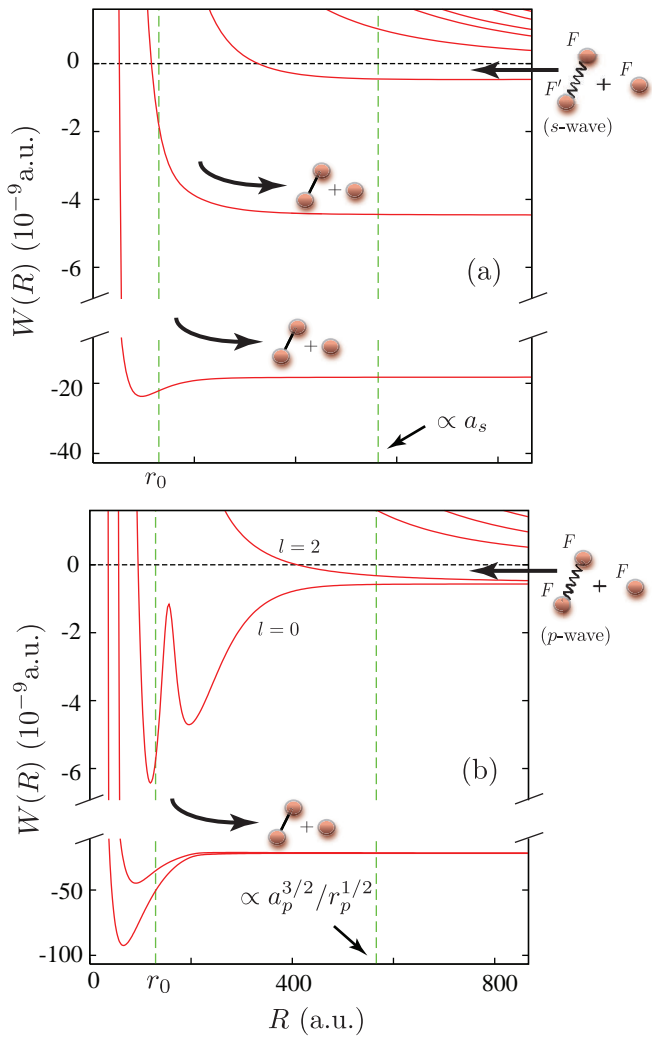}
\caption{(Color online) Three-body effective potentials for (a) $J^\pi=0^+$ $FFF'$ fermionic systems with $s$-wave resonant interaction and for 
(b) $J^\pi=1^-$ $FFF$ spin-polarized fermionic systems with $p$-wave resonant interactions. As is discussed in the text, these potentials 
illustrate the main mechanisms that produce different collisional behaviors of the $s$- and $p$-wave molecules presented here.}
\label{potentials}
\end{figure}

At ultracold energies, the dominant contribution for both $F+FF$ and $F+FF'$ collisions comes from the relative $s$-wave, which 
implies that the dominant three-body symmetry for $FFF'$ is $J^\pi=0^+$ 
($J$ is the total orbital angular momentum and $\pi$ is the total parity) while for $FFF$ it is $J^\pi=1^-$. Figure~\ref{potentials} 
shows typical effective potentials $W_\nu$ for $FFF'$ [Fig.~\ref{potentials}(a)] and $FFF$ systems [Fig.~\ref{potentials}(b)] obtained for
$a_{s}\approx300$ and $a_{p}\approx170$~a.u., respectively. 
For these scattering lengths, the binding energy for $s$-wave molecules, $E^{s}_{b}\approx1/ma^2_s$, and $p$-wave molecules, 
$E^p_{b}\approx2r_{p}/ma_p^3$ \cite{PwaveBindEnergy}, are comparable, making it easier to contrast the two cases.

In Fig.~\ref{potentials} the series of effective potentials that converge to zero as $R\rightarrow\infty$ are three-body continuum channels 
representing collisions between three free atoms. Lower-lying potentials represent atom-molecule entrance or escape channels; notice that there are two
$FF+F$ channels in Fig.~\ref{potentials}(b) converging to the $p$-wave $FF$ molecule energy
at large $R$ due the possible $l=0$ and $l=2$ atom-molecule relative angular momenta allowed for $J^\pi=1^-$. 
In Fig.~\ref{potentials} we also sketch the elastic and inelastic pathways  that are of interest here. 

From the figure, we notice a 
fundamental difference between the initial $FF'+F$ and $FF+F$ collision channels. For $FF'+F$ collisions, the effective potential 
is repulsive (proportional to $1/R^2$ \cite{DIncao}) in the range $r_{0} \ll R\ll a_{s}$, while the $l=0$ effective potential for 
$FF+F$ collisions is attractive in the analogous range $r_{0}\ll R \ll a_{p}^{3/2}/r_{p}^{1/2}$ [the quantity $a_{p}^{3/2}/r_{p}^{1/2}$ 
can be understood from the length scale defined by the molecular binding energy $1/k_{p}=(mE^p_{b})^{-1/2}$]. The 
presence of a repulsive barrier for $FF'+F$ collisions whose range scales with $a_s$ suggests 
that  $a_{ad}$ should be proportional to $a_{s}$, in analogy 
to scattering from a hard sphere of radius $a_{s}$. The inelastic processes for this case, however, should be suppressed as $a_{s}$ increases, 
since it becomes increasingly difficult for the system to tunnel to the region $R\approx r_0$
where the coupling to the deeper molecular states lies [we have verified this statement from our calculations of the nonadiabatic
couplings in Eq.~(\ref{SchrEq})]. In fact, in Ref.~\cite{DIncao} it was demonstrated that the suppression of the inelastic losses
scale with $a$ as $a_{s}^{1-2p_{0}}=a_{s}^{-3.33}$, where $p_{0}$ is related with the strength of the repulsive potential barrier in the range
$r_{0} \ll R\ll a_{s}$.
In contrast, the presence of an attractive
potential for $FF+F$ collisions instead of a repulsive potential allows the atom and molecule to approach each other closely 
without any suppression of the inelastic transition probability. In addition, the absence of a repulsive barrier does not allow 
us to make an analogy to scattering from a hard sphere,
which implies that $a_{ad}$ is not necessarily a simple function proportional to $a_{p}$.

In Refs.~\cite{Castin,Gurarie}, it has been speculated that there is one trimer state in this attractive potential that can cause resonant 
enhancement of the elastic and inelastic rates when varying $a_{p}$. As we show below, in our calculations (see Fig.~\ref{rates}) 
we do observe resonant effects for elastic and inelastic $FF+F$ collisions, and we do associate them with the formation of a trimer state as 
predicted in Refs.~\cite{Castin,Gurarie}. 
However, we believe such trimer states are likely to be non-universal in the sense that their energy as a function of $a_{p}$
depends on the short-range physics not fully represented by the two-body parameters in Ref.~\cite{Castin,Gurarie}. 
Therefore, the resonant peaks in the elastic and inelastic rates will likely depend on a purely three-body non-universal parameter, similar to
the problem with identical bosons where all three-body observables depend on a three-body parameter representing the short-range physics
\cite{Efimov,DIncao,Braaten}.

\section{Elastic and Inelastic atom-molecule collisions}

\begin{figure}[htbp]
\includegraphics[width=3.25in,angle=0,clip=true]{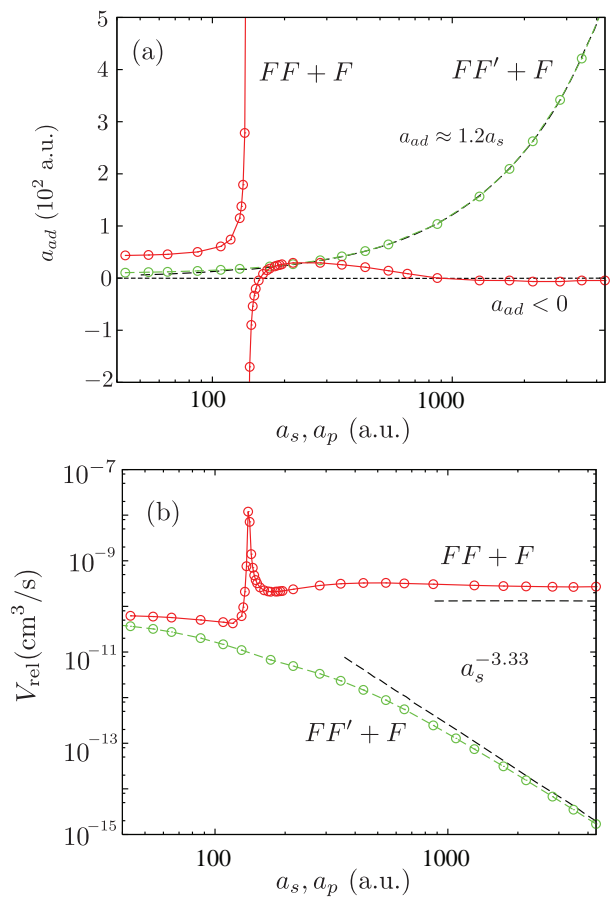}
\caption{(Color online) Numerical calculations for the (a) atom-molecule scattering length $a_{ad}$ and (b) vibrational 
relaxation rate $V_{\rm rel}$ for $J^\pi=0^+$ $FF'+F$ (dash-circle line) and $J^\pi=1^-$ $FF+F$ (solid-circle line) collisions as a function 
of $a_{s}$ and $a_{p}$, respectively. While for $FF'+F$ collisions $a_{ad}$ and $V_{\rm rel}$ show 
a strong dependence on $a_{s}$, for $FF+F$ they are essentially insensitive to variations of $a_{p}$.}
\label{rates}
\end{figure}

We have verified the qualitative behavior described above for the elastic and inelastic rates by
solving the Schr\"odinger equation (\ref{SchrEq}) in coupled hyperradial form using an $R$-matrix approach \cite{Rmatrix}. 
We define the atom-molecule scattering length in terms of the real part of the complex phase-shift obtained from the 
$S$-matrix element [$\exp({2i\delta_{ad}})=S_{ad,ad}$] associated with the atom-molecule channel,
\begin{equation}
a_{ad}=-\lim_{k_{ad}\rightarrow0}\frac{{\rm Re}\left[\tan\delta_{ad}\right]}{k_{ad}}, 
\label{aad}
\end{equation}
\noindent
with $k_{ad}=[2\mu_{ad}(E+E_{b})]^{1/2}$ where $\mu_{ad}=2m/3$ and $E_{b}$ is the binding energy of the weakly bound 
dimer. For inelastic collisions, we define the vibrational relaxation rate $V_{\rm rel}$ as
\begin{equation}
V_{\rm rel}=\sum_{f}\frac{(2J+1)\pi}{\mu_{ad}k_{ad}}|S_{f\leftarrow i}|^2,
\end{equation}
\noindent
where $i$ represents the initial collision channel (atom+weakly bound molecule); and $f$, all possible
final collision channels (atom+deeply bound molecule). 
We have calculated $a_{ad}$ and $V_{\rm rel}$ for values of $a_{s}$ and $a_{p}$ ranging from 
$r_{0}(\approx 65$~a.u.) up to approximately $70r_{0}$ at a collision energy of 1~nK, in order to satisfy the $k_{ad}\rightarrow0$
limit in Eq.~(\ref{aad}). The results are shown in Figs.~\ref{rates}(a) and (b). 

For $FF'+F$ collisions, we have obtained both $a_{ad}$ 
and $V_{\rm rel}$  in the $a_{s}\gg r_{0}$ regime that verifies the $a_{ad}\approx1.2a_{s}$ 
universal prediction~\cite{Petrov} and reproduces the strong $a_{s}^{-3.33}$ suppression
of the inelastic rate~\cite{Petrov,DIncao}.  This scenario is rather favorable for the realization 
of cold atom-molecule mixtures since it combines strong elastic rates with suppression of atomic 
and molecular losses.

On the other hand, $FF+F$ elastic and inelastic processes involving $p$-wave molecules  are quite different.
For smaller values of $a_{p}$, we observe a strong variation of both $a_{ad}$ and $V_{\rm rel}$ [see Fig.~\ref{rates}] due to 
the presence of a trimer state \cite{Castin,Gurarie} that becomes unbound as $a_{p}$ increases. 
As we mentioned before, we do not expect the energy of such a trimer state, nor the value of $a_p$ at which it appears,
to be universal, since they both depend on the short-range physics. Consequently,
the resonance position in Fig.~\ref{rates} can change, and eventually disappear, for different choices of the two-body interaction. 
However, near a $p$-wave two-body Feshbach resonance, i.e., in the regime $a_{p}\gg r_0$, 
our results for $a_{ad}$ and $V_{\rm rel}$ are insensitive to variations of $a_{p}$. 
We rationalize this result by recognizing that even though $a_p$ changes substantially, 
the size of the weakly-bound $p$-wave molecule is practically unchanged.
We expect that this insensitivity of $a_{ad}$ and $V_{\rm rel}$ to variations of $a_{p}$ is universal but
that their actual values will not be universal since the system must access small $R$ where 
details of the two-body interactions become important. 
Further, it can be demonstrated by a simple WKB analysis \cite{DIncao} that the constant character of $V_{\rm rel}$ as a 
function of $a_{p}$ is ensured provided the effective potential falls off faster than $1/R^2$. 
From our numerical calculations, for instance, we found that $a_{ad}<0$ 
in the regime $a_{p}\gg r_{0}$, in contrast to the $a_{ad}>0$ result of Refs.~\cite{Castin,Gurarie}.

The consequences of the three-body physics discussed above for the lifetime and stability of spin-polarized fermionic mixtures of atoms and
molecules are very different from the scenario found
for a gas of fermions in two different spin states. The constant value of $a_{ad}$ for $a_{p}\gg r_{0}$ means that 
atom-molecule collisions will not help much for evaporative cooling, unless $a_{ad}$ is unnaturally large. 
More importantly, the insensitivity of $V_{\rm rel}$ to $a_{p}$ implies that the molecular lifetimes do not change as 
we approach the point of divergence of $a_{p}$, in contrast to $s$-wave molecules whose lifetime increases near a zero-energy Fano-Feshbach resonance. 
These results are in agreement with recent experimental data for an atom-molecule gas mixture of $^{40}$K atoms \cite{PwaveMolecules}, 
where it was found that the lifetimes of both $m_{p}=0$ and $|m_{p}|=1$ molecules do not depend on the magnetic field and therefore on $a_p$. 
In Ref.~\cite{PwaveMolecules}, however, it was found that the molecules were shorter-lived than one would expect based on the magnetic field
independent {\em two}-body dipolar relaxation rates.
More recently \cite{DSJin}, loss rates substantially below those found in \cite{PwaveMolecules} were observed after the atoms had
been removed from the trap, leaving only molecules.
Therefore, since both dipolar relaxation and atom-molecule collisions yield 
molecular losses independent of the magnetic-field, we believe that the shorter molecular lifetimes were due to vibrational relaxation as we
have calculated here.
Our prediction of approximately constant atom-molecule losses also demonstrates that even when atoms are prepared in their 
lowest hyperfine state, and therefore 
dipolar relaxation is energetically forbidden, a magnetic field independent molecular lifetime is likely to be found for other atomic 
species, 
but now limited by three-body loss processes. In the absence of atoms, molecule-molecule elastic and inelastic processes should also be important. 
We speculate that these are also likely to be insensitive to variations of $a_{p}$ (see also Ref.~\cite{Gurarie}), based on the fact that the size of 
a $p$-wave molecule does not depend on $a_{p}$.

\section{Few-body parameters for many-body theories}

The fact that $a_{ad}$ does not seem to depend on $a_p$ raises the question whether $a_{ad}$ will be the important parameter 
for theories attempting to describe many-body aspects of atom-molecule mixtures in spin-polarized Fermi gases. A constant value 
of $a_{ad}$ across a Feshbach resonance essentially means that the atom-molecule interaction does not ``feel'' the presence of the $p$-wave 
resonance. 
From this perspective, the atom-molecule interaction is not controllable, and 
the atom-molecule mixture cannot be claimed to be strongly interacting near the resonance.
For this reason, we have also performed some preliminary calculations to explore $p$-wave  atom-molecule collisions. 
For $J^{\pi}=0^+$, $1^+$, and $2^+$ (which are all the symmetries that allow $p$-wave collisions between an atom and a $p$-wave molecule),
we have found that the three-body effective potentials are now repulsive in the range $r_{0}\ll R\ll a_{p}^{3/2}/r_{p}^{1/2}$, instead 
of attractive as found for $J^{\pi}=1^-$. The presence of such a repulsive barrier suppresses inelastic $p$-wave atom-molecule collisions 
\cite{DIncao}, making these collisions less important than $s$-wave collisions in determining the lifetime of $p$-wave molecules. 
On the other hand, the repulsive barrier indicates 
that the $p$-wave atom-molecule scattering length, $a^p_{ad}$, should be proportional to
$a_{p}^{3/2}/r_{p}^{1/2}$, again in analogy to scattering from a hard-sphere.
Therefore, it is likely that $a^p_{ad}$ will be an important parameter for many-body theories of atom-molecule mixtures near a 
$p$-wave resonance, since $a^p_{ad}$ is a controllable parameter. 
Experimentally, however, $p$-wave atom-molecule interactions will be comparably difficult to control as $p$-wave atom-atom 
interactions because of the suppression of the energy dependent elastic cross-sections. 
Certainly, it will be necessary to explore in depth the importance of $a^p_{ad}$, but this discussion 
is outside the scope of this paper. 

\section{Summary}

In this paper, we have determined the atom-molecule scattering length, $a_{ad}$, and vibrational relaxation rate, 
$V_{\rm rel}$, for collisions involving $s$- and $p$-wave molecules of fermionic atoms as a function of $a_{s}$ and
$a_{p}$ which can be controlled with two-body Feshbach resonances. 
We have determined $a_{ad}$ and $V_{\rm rel}$ for $s$-wave molecules of 
fermionic atoms in different spin states, verifying previously derived results. We could thus identify the 
mechanisms that control the collisional properties of $p$-wave molecules. For spin-polarized systems,
we have found that both $a_{ad}$ and $V_{\rm rel}$ are insensitive to variations of the two-body $p$-wave scattering length $a_{p}$ 
for $a_{p}\gg r_{0}$, consistent with the recent experimental data in Ref.~\cite{PwaveMolecules}. We expect the 
results for $a_{ad}$ and $V_{\rm rel}$ to be universal with respect to their insensitivity to variations of $a_{p}$ 
but their values can change for different atomic species. We also speculated that, due to the constant character of $a_{ad}$,
the relevant parameter for many-body theories for a spin-polarized mixture of atoms and molecules is likely to be associated with
$p$-wave collisions between a fermionic atom and a $p$-wave molecule. Our preliminary results indicate that the $p$-wave
scattering length for such collisions, $a^{p}_{ad}$, is likely to be proportional to $a_{p}^{3/2}/r_{p}^{1/2}$.

Note that the very recent preprint by Jona-Lasinio {\em et al.} \cite{Castin} argues that the threshold exponent for three-body 
recombination of spin-polarized fermions should be proportional to $K_3 \propto a_p^{15/2} r_p^{1/2}$ at large $p$-wave scattering lengths. 
This differs slightly from the prediction of Suno {\em et al.} \cite{OurFermion} that $ K_3 \propto a_p^8 $. The derivation of 
Suno {\em et al.} was, in fact, based on dimensional analysis, starting from the assumption that $a_p$ was the dominant length 
scale.
In fact, it is now understood that $a_p$ alone does not adequately describe the dominant length scale for two-body $p$-wave
scattering, implying that a modification of the dimensional analysis prediction of Suno {\em et al.} should be expected.
Investigation of this point will be relegated to future studies.

This work was supported by the National Science Foundation and the Air Force Office of Scientific Research. 
We also want acknowledge the Keck Foundation for providing computational resources.


\begin{thebibliography}{99}
\bibitem{PwaveBindEnergy} This approximate value for the binding energy of $p$-wave dimers
can be obtained assuming the effective range expansion $k^{3}\cot{\delta_{p}}\approx-1/a_{p}^3-k^2/2r_p$
and determining the poles of the scattering amplitude.
\end{thebibliography}
\end{document}